\documentclass[aps,12pt, final, notitlepage, oneside,
onecolumn, nobibnotes, nofootinbib, superscriptaddress, noshowpacs, centertags]{article}
\usepackage{graphicx}
\begin{document}

\title{LOW METALLICITY SOLAR MODEL FINDS THE PROVE IN BOREXINO DATA}
\author{L.~B.~Bezrukov$^1$, A.~S. Bykovsky$^{1,2}$, V.~V.~Sinev$^{1,2}$ \\
$^1$ Institute for Nuclear Research, Moscow, 117312  Russia\\
$^2$ National Research Nuclear University, Moscow, Russia}


\maketitle
\begin{abstract}
 The Borexino Collaboration interprets its data within the framework of Bulk Silicate Earth model and the model of the Sun with high metallicity. Other authors have given a different interpretation of the same data within the framework of Hydridic Earth model and low-metallicity Sun model. In order to understand what occasion takes place in Nature, the Borexino single events energy spectrum was numerically simulated using the Monte Carlo method for various assumptions about the processes that could exist in Nature. The existence of large potassium geo-antineutrino flux and its absence were considered. At the same time, the high or low metallicity of the Sun were included in simulations. A comparison of counting rates reconstruction from simulated data with the reconstruction ones from Borexino single events spectrum demonstrates that large potassium geo-antineutrino flux with low metallicity of the Sun is preferable and is realized in Nature.

\vspace{5mm}

\end{abstract}

\section{Introduction}

The Borexino Phase III  \cite{borex23} data are analyzed in  \cite{bezruk24} upon adding known antineutrino fluxes to the backgrounds and solar neutrino fluxes. This analysis made it possible to determine the flux of $^{40}$K geo-antineutrinos, which turned out to be quite significant in magnitude. At the same time, all fluxes of solar neutrinos are compatible with the low-metallicity model of the Sun. 
To make a comprehensive analysis of this result we performed numerical simulations of the Borexino experiment spectrum as in \cite{bezruk23} using various assumptions about the processes possibly existing in Nature. 

We will use the following assumptions about Nature: Bulk Silicate Earth (BSE) model or Hydridic Earth (HE) model \cite{Thoulhoat22}, high-metallicity (HZ) or low-metallicity (LZ) solar model. So, we have four assumptions: BSE+LZ, BSE+HZ, HE+LZ, HE+HZ. Which of these assumptions has higher probability to be realized in Nature?

BSE and HE models have different predictions, in particular, on potassium abundance in the Earth's interior. The BSE model predicts the 40K geo-antineutrino events counting rate in Borexino scintillator  $R(^{40}\rm K)$ close to zero but the HE model predicts it to be large, up to  $R(^{40}\rm K) = 10$ cpd/100t. Below for the $R$ we will use the following units: cpd/100t (counts per day per 100 ton of scintillator).

We perform two types of multivariate fit analyses here: with $^{40}$K geo-antineutrino event PDF (Probability Density Function) and without it. The second type of the analysis was performed in Borexino Collaboration. So, we may have up to 8 scenarios of different analyzes of simulated pseudo-experiments: “BSE+HZ+fit without 40K”, “BSE+HZ+fit with 40K”, “BSE+LZ+fit without 40K”, “BSE+LZ+fit with 40K”, “HE+HZ+fit without 40K”, “HE+HZ+fit with 40K”, “HE+LZ+fit without 40K”, “HE+LZ+fit with 40K”. After making the simulations according to pointed up conditions we compare each analysis with ones performed earlier using the available experimental data sample by the Borexino Collaboration: the original Borexino Collaboration analysis  \cite{borex23} and the analysis performed in the work  \cite{bezruk24}.

\section{RESULTS OF EXPERIMENTAL BOREXINO DATA SAMPLE FITS}

     In order to make it easier to compare the results of the analyzes of simulated data with experimental ones, we present here the results of the Borexino Collaboration analysis \cite{borex23} and the analysis from \cite{bezruk24} which are shown in Table \ref{tabl:cnoflux}. 

In the Table \ref{tabl:cnoflux} the Borexino Phase-III results on counting rates $R$(7Be (862 + 384 keV)), $R$($pep$), $R$(8B), $R$(CNO) in cpd/100t units are shown in the first column. In the second one we show results of our analysis without including $^{40}$K geo-anineutrinos according to the receit of the Borexino Collaboration. It is clearly seen that results are the same. 

Then we have included in our analysis events from $^{40}$K geo-antineutrinos. The result is presented in the column 3 where we used fixed value for counting rate of potassium events $R(^{40}\rm K) = 11$ cpd/100t in multivariate fit analysis. In the column 4 we left free the parameter of  $R(^{40}\rm K)$.  It is appeared that its value reached 19 cpd/100t, but counting rate of CNO events fall dawn to 2.4 cpd/100t, until the limitation we have installed for the CNO events. In parallel the $\chi{^2}$ value comes to the smaller value an satisfies better to the statistical behavior. The columns 5 and 6 labeled as HZ and LZ show the theoretical interaction rates predicted by the standard solar model under the high (HZ) and low (LZ) metallicity \cite{borex19}. 

Regarding the Table \ref{tabl:cnoflux} one can conclude that the result presented in 4th column is preferable in comparison with columns 2 and 3.

\begin{table}[ht]
\caption{Counting rates $R$(7Be), $R$($pep$), $R$(8B), $R$(CNO) and $R$(40K geo-$\nu$) in cpd/100t units following from Borexino Phase-III data with our analysis. The uncertainties for counting rates presented here are ones following from the minimizing $\chi{^2}$ function by the ROOT code during multivariate fit analysis. The values shown without errors were fixed during analyses. The last row shows the  $\chi{^2}$ values obtained in multivariate fit analysis. In column 1 the Borexino Collaboration results are shown. The column 2 shows the result of our analysess reproducing values presented by the Borexino Collaboration. The column 3 shows the result of addition the ${^40}$K PDF with fixed $R$(40K) = 11 value. The column 4 shows the result with free parameter of $R$(40K). The columns 5 and 6 labeled as HZ and LZ show the theoretical interaction rates predicted by the standard solar model under the high (HZ) and low (LZ) metallicity \cite{borex19}.} 
\label{tabl:cnoflux}
\centering
\vspace{2mm}
\begin{tabular}{| c | c | c | c | c | c | c |}
\hline
\hline
&1&2&3&4&5&6 \\
\hline
&Borexino&Borexino&+ fixed   &Analysis with  & HZ \cite{borex19}&LZ  \cite{borex19} \\ 
&            & like      & $R$(40K)&free $R$(40K)& &\\
\hline
$R$(7Be)&48.3$\pm$2.5&48.4$\pm$1.2&45.9$\pm$1.3&43.6$\pm$1.5&47.9$\pm$2.8&43.7$\pm$2.5\\
\hline
$R$($pep$)&2.74$\pm$0.04&2.74&2.74&2.9$\pm$0.3&2.74$\pm$0.04&2.78$\pm$0.04\\
\hline
$R$(8B)&0.2$\pm$0.1&0.16&0.16&0.14$\pm$0.05&&\\
\hline
$R$(CNO)   &6.7$\pm$1.2&7.6$\pm$1.2&4.4$\pm$0.6&2.6$\pm$0.6&4.92$\pm$0.55&3.52$\pm$0.37\\
\hline
$R$(40K)    &0&0    &11&19.1$\pm$2.5&&\\
\hline
$\chi{^2}$ & &198 &171 &161 & & \\  
\hline
\hline
\end{tabular}	
\end{table}

\begin{figure*}[t!]
\begin{center}
\includegraphics[width=180mm]{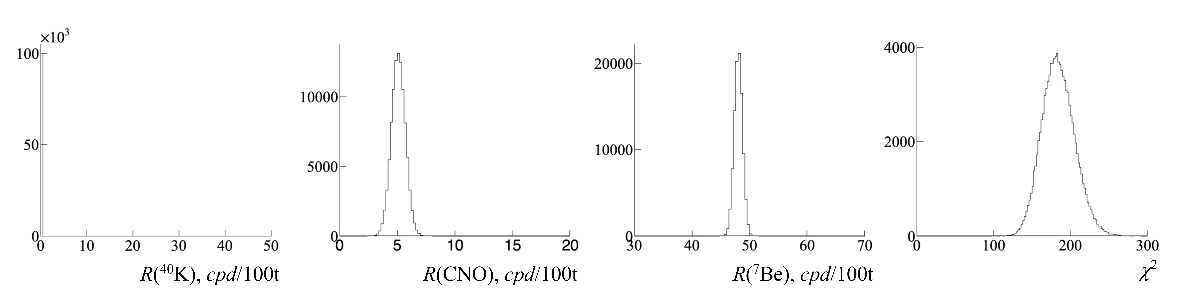}
\end{center}
\caption{\label{figone} Distributions of counting rates $R$ and $\chi{^2}$  found by multivariate fit of simulated Borexino data. The values on vertical axes are given in relative units. The data simulated under assumptions: $\bar R$ (40K) = 0; the Sun has HZ metallicity:  $\bar R$(CNO) = 5,  $\bar R(pep)$ = 2.74,  $\bar R$(7Be) = 48. During the fit we followed to the Borexino Collaboration and used $R$(40K) = 0 fixed. The obtained mean counting rates are: $<R(\rm CNO)>$ = 5,1, $<R(pep)>$ = 2.747, $<R(7\rm Be)>$ = 48. The obtained mean value of $\chi{^2}$ =184.6.} 
\end{figure*}

\begin{figure*}[t!]
\begin{center}
\includegraphics[width=180mm]{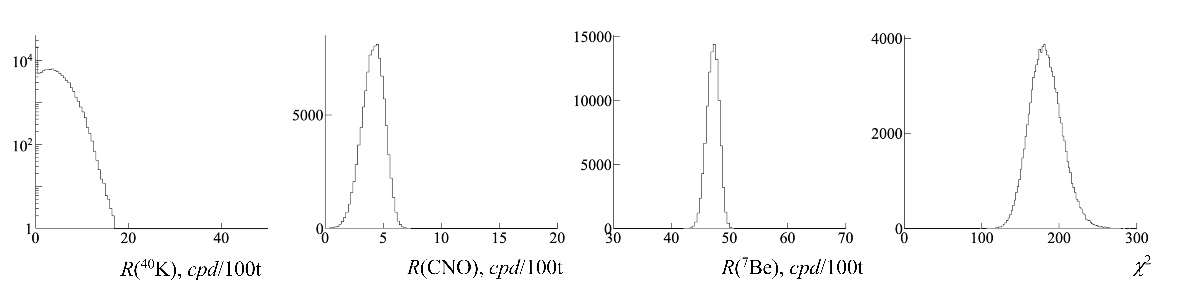}
\end{center}
\caption{\label{fig2} Distributions of counting rates $R$ and $\chi{^2}$ found by multivariate fit of simulated Borexino data. The values on vertical axes are given in relative units. The data simulated under the same assumptions as shown at Fig. 1. $R$(40K) is free in the fit. The obtained mean counting rates are: $<R(40\rm K)>$ = 3.42, $<R(\rm CNO)>$ = 4.11, $<R(pep)>$= 2.77, $<R(7\rm Be)>$ = 47.1. The obtained mean value of $\chi{^2}$ =182.8.}
\end{figure*}

\begin{figure*}[t!]
\begin{center}
\includegraphics[width=180mm]{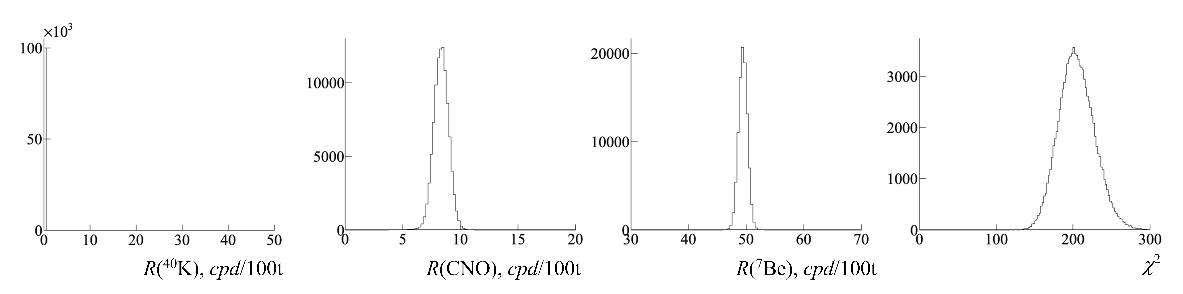}
\end{center}
\caption{\label{fig3} Distributions of counting rates $R$ and $\chi{^2}$ found by multivariate fit of simulated Borexino data. The values on vertical axes are given in relative units. The data simulated under assumptions: $\bar R (40\rm K) $= 14; the Sun has LZ metallicity: $\bar R(\rm CNO)$ = 3.5, $\bar R (pep)$ = 2.9, $\bar R(7\rm Be)$ = 46. In the fit we followed to the Borexino Collaboration and used $R$(40K) = 0 fixed. The obtained mean counting rates are: $<R(40\rm K)$ = 0, $<R(\rm CNO)>$ = 8.4, $<R(pep)>$ = 2.5, $<R (7\rm Be)>$ = 49.2. The obtained mean value of $\chi{^2}$ =205.  }
\end{figure*}

\begin{figure}[ht]
\begin{center}
\includegraphics[width=180mm]{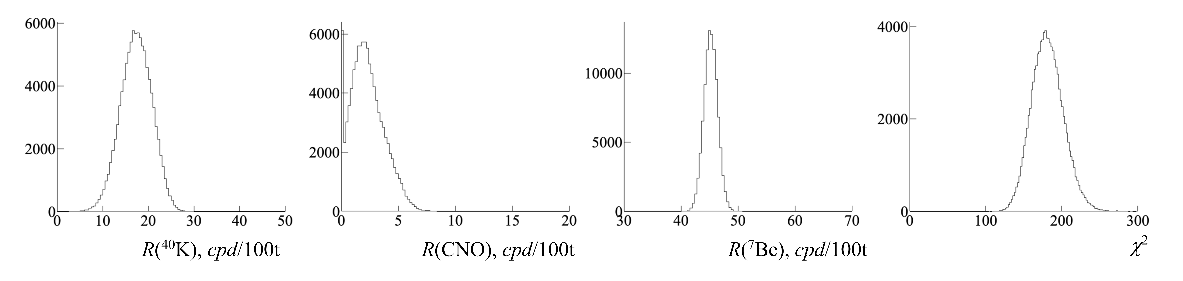}
\end{center}
\caption{\label{fig4} Distributions of counting rates $R$ and $\chi{^2}$ found by multivariate fit of simulated Borexino data. The values on vertical axes are given in relative units. The data simulated under the same assumptions as shown at Fig. 3. In the fit here we used $R$(40K) free opposite to the analysis at Fig. 3. The obtained mean counting rates are: $<R(40\rm K)>$ = 18, $<R(\rm CNO)>$ =1.53, $<R(pep)>$ = 3.3, $<R(7\rm Be)>$ = 45,3. The obtained mean value of $\chi{^2}$ =181.}
\end{figure}

\begin{figure*}[t!]
\begin{center}
\includegraphics[width=180mm]{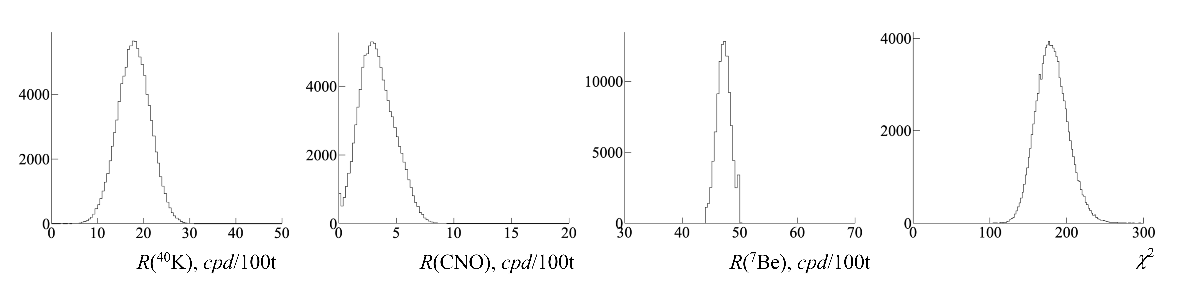}
\end{center}
\caption{\label{fig5} Distributions of counting rates $R$ and $\chi{^2}$ found by multivariate fit of simulated Borexino data. The values on vertical axes are given in relative units. The data simulated under assumptions that $\bar R$(40K) = 14 and Sun has HZ metallicity: $\bar R$(CNO) = 5, $\bar R(pep)$ = 2.74, $\bar R$(7Be) = 48. During the fit here we left $R$(40K) free. The obtained mean counting rates are:$<R(40\rm K)>$ = 17.1, $<R(\rm CNO)>$ = 4.09, $<R(pep)>$ = 2.77, $<R(7\rm Be)>$ = 47.2. The obtained $\chi{^2}$ mean value 181.8.}
\end{figure*}

\section{RESULTS OF NUMERICAL PSEUDO EXPERIMENTS}

We have simulated the single events Borexino experimental spectrum as in \cite{bezruk23} for certain set of processes under assumptions on the mean counting rate $\bar R$ for each process. All PDFs of solar neutrinos, 40K geo-antineutrino and backgrounds were included in simulation. 
When preparing the simulated spectrum, we made our analysis for reconstructing by multivariate fit the counting rates $R$ only for those processes that one can expect as existed in the Nature. For simplicity we fixed all backgrounds during fit. For each scenario we have simulated $10^5$ spectra corresponding to the Borexino one.

At Figure \ref{figone} the results of simulating “BSE+HZ+fit without 40K” are presented. The data were simulated here under assumptions that $\bar R$(40K) = 0 and the Sun has HZ metallicity: $\bar R$(CNO) = 5, $\bar R(pep)$ = 2.74, $\bar R$(7Be) = 48. These are generally accepted assumptions. The obtained mean counting rates after $10^5$ spectra analysis are: $<R(\rm CNO)>$ = 5.1, $<R(pep)>$ = 2.747, $<R(7\rm Be)>$ = 48. The obtained mean value of $\chi{^2}$ =184.6. For 162 experimental points obtained value of $\chi{^2}$ =184.6 is rather large. The reason of this is that we fixed counting rate values of some processes during fit. During the fit we used the conditions like this: $R$(40K) = 0, $R$(8B) were fixed, counting rate of $pep$ reaction was constrained as $R(pep)$ = 2.74 ± 0.04 and counting rates of $R$(CNO) and $R$(7Be) were left free. So, for the absence of 40K events, as it was expected, there is no 40K counting rate distribution. CNO counting rate event distribution groups near the simulated value (5 cpd/100t) and 7Be also places around simulated value 48 cpd/100t. Let’s note that $\chi{^2}$ mean value is slightly more than 180. Let’s fix this observation. 

So, we can conclude that if there is no potassium and the HZ model is valid, the CNO events are good reconstructed and $\chi{^2}$ would be significantly less than 200. Let’s compare our fit result with the result in second column of Table \ref{tabl:cnoflux}. We can find that Borexno Collaboration result is differ from obtained here: Borexino $R$(CNO) and $\chi{^2}$ are higher than must be for the scenario “BSE+HZ+fit without 40K”. So, the interpretation commonly accepted, that Borexino data support the BSE model and HZ solar model is incorrect. 

At Figure \ref{fig2} we present the result of the same simulation as presented above but the analysis was made with looking for 40K events. We call it “BSE+HZ+fit with 40K” scenario. The obtained mean counting rates appeared like this: $<R(40\rm K)>$ = 3.42, $<R(\rm CNO)>$ = 4.11, $<R(pep)>$ = 2.77, $<R(7\rm Be)>$ = 47.1. The obtained mean value of distribution for $\chi{^2}$ values is 182.8. We have found small amount of 40K antineutrino events due to statistical fluctuations of other components. The systematic bias for the average values of $<R(\rm CNO)>$ and $<R(7\rm Be)>$ was obtained. Regarding these results we can conclude that in absence of 40K flux in the Nature it could be found in small amount as a result of statistical fluctuations. The $\chi{^2}$ value comes to be slightly better than in scenario of neglecting this flux in an analysis, but stays very close to the first scenario. 

Third scenario of simulation and analysis is shown at Figure \ref{fig3}. We call it “HE+LZ+fit without 40K and it reflects the scenario of LZ solar model and large potassium abundance in the Earth. For the simulation we used mean counting rate like this: $\bar R$(40K) = 14, $\bar R$(CNO) = 3.5, $\bar R(pep)$ = 2.9, $\bar R$(7Be) = 46. In the fit we did not include the 40K geo-antineutrinos following the Borexino Collaboration. The obtained mean counting rates are: $<R(\rm CNO)>$ = 8.4, $<R(pep)>$ = 2.5, $<R(7\rm Be)>$ = 49.4 that corresponds to the Borexino Collaboration result \cite{borex23} and to the values in first and second columns of Table \ref{tabl:cnoflux}. The obtained mean value of $\chi{^2}$ distribution is 205 what coincides rather good with values $\chi{^2}$ = 198 from the Table \ref{tabl:cnoflux}. The comparison of this set of parameters for the simulation with the reconstructed ones demonstrates that result of the Borexino Collaboration could be interpreted as in the Nature can be with high probability realized the scenario of large potassium abundance in the Earth and simultaneously the validity of LZ solar model. 

At  Figure \ref{fig4} we demonstrate the distributions of counting rates $R$ and $\chi{^2}$ found by multivariate fit of simulated Borexino data for the scenario “HE+LZ+fit with 40K”. We used the same assumptions for the simulation as in previous case: $\bar R$(40K) = 14, LZ metallicity with $\bar R$(CNO) = 3.5, $\bar R(pep)$ = 2.9 and $\bar R$(7Be) = 46. During the fit here we used $R$(40K) free opposite shown at Figure \ref{fig3} analysis. We obtain the following mean counting rates: $<R(40\rm K)>$ = 18, $<R(\rm CNO)>$ = 1.53, $<R(pep)>$ = 3.3, $<R(7\rm Be)>$ = 45.3. The mean value of $\chi{^2}$ distribution appear to be 181. This result corresponds to our analysis of the Borexino experimental sample of data shown in column 4 of Table \ref{tabl:cnoflux}. This scenario also demonstrates that results of the Borexino Collaboration could be interpreted the same way as in scenario 3 shown above. The large potassium abundance in the Earth and LZ solar model can be realized in the Nature with high probability. It is interesting that this conclusion coincides with one we made for the analysis shown at  Figure \ref{fig3}. 

We perform the simulation for the scenario “HE+HZ+fit with 40K” at figure  \ref{fig5}. The data were simulated under assumptions that counting rate of 40K is high ($\bar R$(40K) = 14) and the Sun has HZ metallicity: $\bar R$(CNO) = 5, $\bar R(pep)$ = 2.74, $\bar R$(7Be) = 48. $R$(40K) was left free in the fit. The obtained mean counting rates are: $<R(40\rm K)>$ = 17.1, $<R(\rm CNO)>$ = 4.09, $<R(pep)>$ = 2.77, $<R(7\rm Be)>$ = 47.2. The obtained mean value of $\chi{^2}$ distribution is 181.8. This result is worse than our analysis of the Borexino experimental data sample for the scenario of low metallicity shown at Figure \ref{fig4}. This simulation demonstrates that the assumption about LZ is more preferable. 

We do not perform here the results of other scenarios “BSE+LZ+fit with 40K”, “BSE+LZ+fit without 40K”and “HE+HZ+fit without 40K” because they sufficiently differ from results of both available analyzes of experimental data sample.

\section{DISCUSSION}

The possible potassium abundance in the Earth was discussed in \cite{bezruk23} based on value $R$(40K) = 11 cpd/100t. The column 3 of Table 1 shows this case. The systematic bias observed in pseudo experiments was found as 4 cpd/100t. So, the value of possible potassium abundance equals to (11 – 4)/2.18 = 3.2\% of the Earth mass was discussed in \cite{bezruk23}. Coefficient 2.18 was calculated in \cite{bezruk23} for 1\% abundance and uniform potassium distribution in the Earth. 

The value $R$(40K) = 19 cpd/100t shown in the column 4 of Table 1 we and \cite{bezruk24} introduce as the result of experimental data analysis following the idea of minimal $\chi{^2}$ value. Taken into account the systimatic bias and distributon of $R$(40K) at Fig. 4 we can obtain the value of the possible potassium abundance in the Earth in the range (5 ÷ 7.8)\% of Earth mass for the case of uniform potassium distribution in the Earth. If the distribution is not uniform, then the potassium abundance may be less. A precise determination of the potassium abundance in the Earth requires a new experiment.

\section{CONCLUSION}

The result of our simulation shows that the assumption HE+LZ is realized in Nature. One can conclude it when compares the simulated scenario “HE+LZ+fit with 40K” with the analysis of Borexino data experimental sample made in \cite{bezruk24} and reproduced by us. Numerical experiment confirms the analysis result. 

 In the same time scenario “HE+LZ+fit without 40K” satisfies to the analysis of Borexino Collaboration \cite{borex23}. And again, numerical experiment confirms the analysis result by Borexino Collaboration. 
Both scenarios included 40K antineutrinos in simulation. This means that both analyses, our and Borexino Collaboration, are correct and they prove the existing of large potassium abundance in the Earth. 
The result of our simulation shows that other assumptions (HE+HZ, BSE+HZ, BSE+LZ) cannot be realized in Nature because all scenarios for these assumptions do not satisfy to both available analyzes of experimental data sample \cite{borex23}, \cite{bezruk24}.

\vspace{2mm}
\section*{Acknowledgments}

Authors are grateful to organizing committee of Session-Conference of the Nuclear Physics Section of the Department of Physical Sciences of Russian Academy of Sciences "Physics of Fundamental Interactions", dedicated to the 70th anniversary of Valery Anatolyevich Rubakov (17 - 21 February 2025, Moscow), for the opportunity to present a report on the topic of this article.

\vspace{2mm}

\end{document}